\documentclass[10pt,aps,floatfix,prd,twocolumn]{revtex4-2}
\usepackage{acronym,dcolumn,graphicx,hyperref,textcomp}
\begin{document}

\title{Multi-messenger detectability of continuous gravitational waves from the
near future to next generation detectors}
\author{Benjamin J. Owen}
\affiliation{
  Department of Physics and Astronomy,
  Texas Tech University,
  Lubbock, Texas 79409-1051,
  USA
}
\affiliation{
  Department of Physics,
  University of Maryland Baltimore County,
  1000 Hilltop Circle,
  Baltimore, Maryland 21250,
  USA
}
\author{Binod Rajbhandari}
\affiliation{
  Department of Mathematics,
  Texas Tech University,
  Lubbock, Texas 79409-1042,
  USA
}
\affiliation{
  Department of Physics and Astronomy,
  Texas Tech University,
  Lubbock, Texas 79409-1051,
  USA
}
\affiliation{
School of Mathematics and Statistics and Center for Computational Relativity and
Gravitation, Rochester Institute of Technology, Rochester, NY 14623, USA
}
\affiliation{
  Department of Physics,
  University of Maryland Baltimore County,
  1000 Hilltop Circle,
  Baltimore, Maryland 21250,
  USA
}

\begin{abstract}
Continuous gravitational waves have the potential to transform gravitational
wave astronomy and yield fresh insights into astrophysics, nuclear and particle
physics, and condensed matter physics.
We evaluate their detectability by combining various theoretical and
observational arguments from the literature and systematically applying those
arguments to known astronomical objects and future gravitational wave detectors.
We detail and update previous estimates made in support of Cosmic Explorer [M.
Evans \textit{et al.}, arXiv:2306.13745; I. Gupta \textit{et al.}, Class.\
Quantum Grav.\ \textbf{41}, 245001 (2024)].
It is commonly argued that the spins of accreting neutron stars are regulated by
gravitational wave emission and that millisecond pulsars contain a young
pulsar's magnetic field buried under accreted material.
If either of these arguments holds, the first detection of continuous
gravitational waves is likely with near future upgrades of current detectors,
and many detections are likely with next generation detectors such as Cosmic
Explorer and the Einstein Telescope.
A lack of detections in the next several years would begin to raise serious
doubts about current theories of millisecond pulsar formation.
\end{abstract}

\maketitle

\acrodef{ATNF}{Australia Telescope National Facility}
\acrodef{AXIS}{Advanced X-ray Imaging Satellite}
\acrodef{EM}{electromagnetic}
\acrodef{GW}{gravitational wave}
\acrodef{LMXB}{low mass x-ray binary}
\acrodefplural{LMXB}{low mass x-ray binaries}
\acrodef{LVK}{LIGO-Virgo-KAGRA}
\acrodef{ngVLA}{next generation Very Large Array}
\acrodef{SGR}{soft gamma repeater}
\acrodef{SKA}{Square Kilometre Array}

\section{Introduction}

The first \ac{GW} detection of a binary neutron star merger,
GW170817~\cite{GW170817}, with its accompanying \ac{EM} detections began the era
of multi-messenger \ac{GW} astronomy~\cite{GW170817MMA}.
Via tidal effects on the phase of the late stages of the \ac{GW} signal,
GW170817 set the first \ac{GW} constraints on the neutron star equation of
state~\cite{GW170817props, GW170817EOS}.
Those constraints improved when combined with constraints from \ac{EM} astronomy
and other \ac{GW} observations, summarized \textit{e.g.} in~\cite{Li2021,
Yunes2022}.
Combined, the \ac{GW} and \ac{EM} signals also yielded a trove of information on
relativistic jets, astrophysical populations, nuclear physics, particle physics,
general relativity, and cosmology~\cite{, GW170817, GW170817props, GW170817MMA,
GW170817EOS, GW170817Hubble}.
Future multi-messenger detections of neutron stars will tell us more about the
astrophysics and fundamental physics of these extreme objects, especially
through detections of sources other than compact binary
mergers~\cite{Glampedakis2018, Branchesi2023, Evans2023, Corsi2024, Gupta2024,
OwenRMP}.

Here we show that some of these novel signals, continuous \acp{GW} from spinning
deformed neutron stars, are likely to be detected by improved versions of the
``Advanced'' \ac{GW} detectors currently operated by the \ac{LVK}
Collaboration~\cite{AdvancedLIGO, AVirgo, KAGRA}---if currently common beliefs
about millisecond pulsar formation are true.
We also detail and extend arguments~\cite{Evans2023, Gupta2024} reviewed
in~\cite{Corsi2024, OwenRMP} that next generation \ac{GW} detectors such as
Cosmic Explorer~\cite{Evans2021} and the Einstein Telescope~\cite{Hild2011} are
likely to detect many continuous wave signals, especially when working in
concert with next generation \ac{EM} detectors such as the radio Square
Kilometre Array~\cite{Dewdney2009}, Next Generation Very Large
Array~\cite{Wilner2024}, Five-hundred-meter Aperture Spherical
Telescope~\cite{FAST}, and the \ac{AXIS} mission~\cite{Reynolds2023}.
These estimates are more detailed than and complementary to those made in papers
supporting the Einstein Telescope such as \citet{Branchesi2023} and
\citet{ETScience}.

For practical purposes, searches for continuous \acp{GW} can be divided into
four types based on prior knowledge of the target from \ac{EM}
astronomy~\cite{Owen2009}:
(1) \textit{Targeted searches} for known pulsars use full (or nearly full)
timing solutions from \ac{EM} astronomy to precisely target either a signal
waveform with a particular set of parameters or a small region of parameter
space.
A recent set of fully targeted pulsar searches is \cite{O4aPulsars} and a recent
example of wider band partially targeted searches is \cite{O4abNarrowband}.
(2) \textit{Directed searches} for known non-accreting neutron stars without
detected \ac{EM} pulsations need to search a broader band of frequencies and
related parameters.
Recent examples of such searches are \citet{Ming} and \citet{O4aSNRs}.
(3) Directed searches for known non-pulsing accreting neutron stars target the
population known to astronomers as the $Z$-source population of \acp{LMXB}.
This population has more parameter uncertainties, making searches harder, but
has excellent arguments for \ac{GW} emission.
A good example is Sco~X-1~\cite{O3ScoX1, O4aScoX1}.
(4) \textit{All-sky surveys} search a huge parameter space, at great
computational cost, for the many yet unseen neutron stars in the galaxy.
Recent examples without and with possible binary parameters (yet another
complication) are \cite{O3AllSky, O4aAllSky} and \cite{Covas2022,
O4aAllSkyBinary}.
Below we discuss detectability estimates for all these types of searches except
the last, which is not (directly) multi-messenger and is well described
elsewhere~\cite[e.g]{Reed2021, Pagliaro2023, ETScience}.
We focus on multi-messenger searches for objects which are already known from
\ac{EM} detections.

Below we update and expand on the arguments made in \citet{Evans2023} and
\citet{Gupta2024} and summarized in \citet{Corsi2024} and \citet{OwenRMP} that
the first continuous \ac{GW} detection is not just possible but likely within
the next few years, and that by the time of Cosmic Explorer and the Einstein
Telescope there should be many detections.
In fact, by then an absence of continuous \ac{GW} detections would call into
question the standard theory of millisecond pulsar
evolution~\cite[e.g.]{Bhattacharya1991} and the assumption that the spins of
accreting neutron stars are regulated by gravitational wave
emission~\cite{Papaloizou1978b, Wagoner1984, Bildsten1998, Andersson1999b}.
Essentially the field is moving from an era when the first detection
\textit{could} happen (e.g., some neutron star might happen to have a large
deformation due to quirks of its history) to when it \textit{should} happen
(e.g., internal magnetic fields in neutron stars generically produce
deformations).

\begin{figure}
\includegraphics[width=0.45\textwidth]{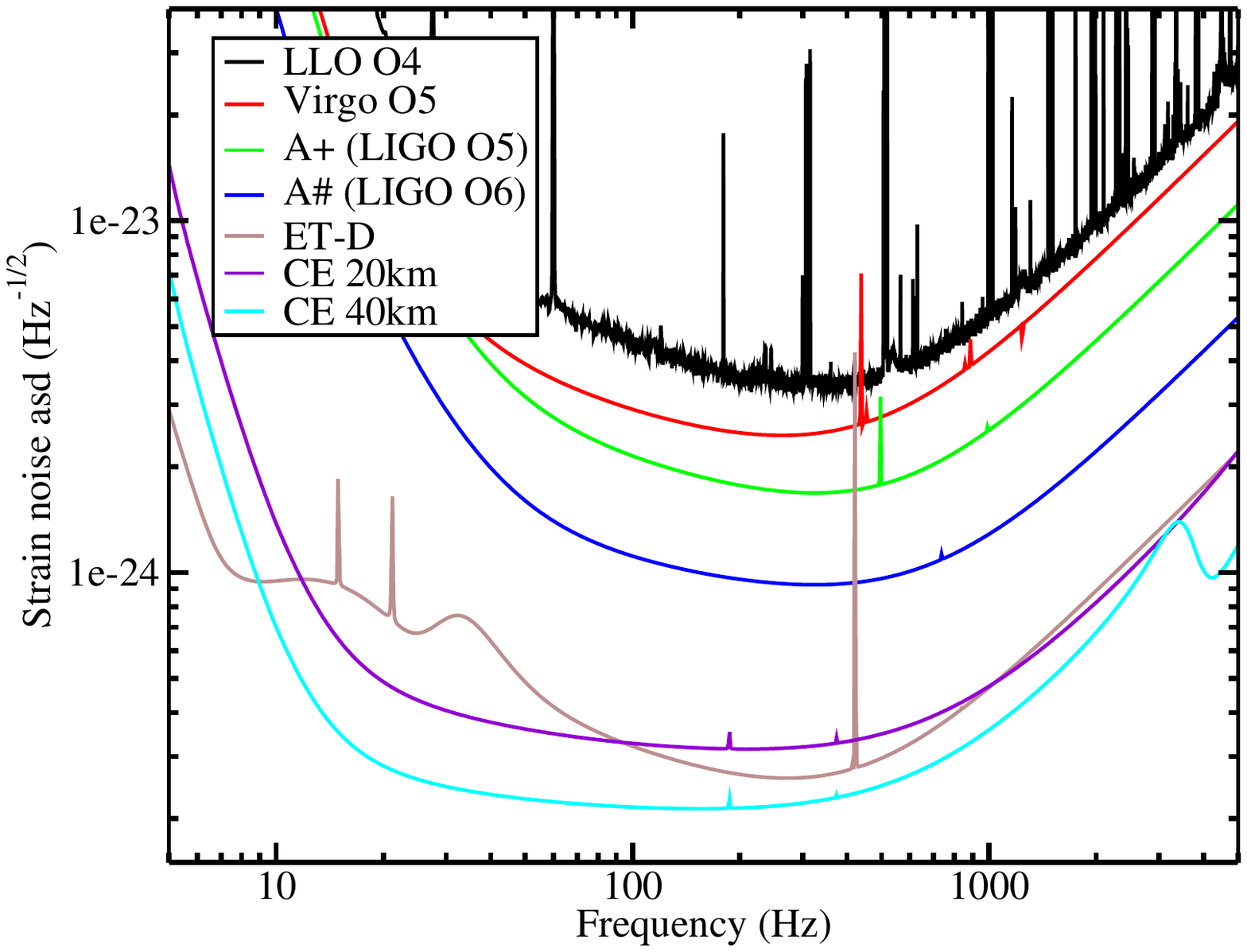}
\caption{
\label{f:noise}
Strain noise amplitude spectral densities (asds) as functions of frequency for a
variety of planned detectors, plus an O4 noise curve from the LIGO Livingston
Observatory (LLO)~\protect\cite{Capote2025} for reference.
}
\end{figure}

We make estimates for a variety of future ground-based detectors.
The \ac{LVK} Collaboration has upgraded the LIGO~\cite{AdvancedLIGO},
Virgo~\cite{AVirgo}, and KAGRA~\cite{KAGRA} detectors to their ``Advanced''
design sensitivity.
Their latest completed joint observing run, the fourth one with Advanced
detectors, is abbreviated O4.
For LIGO the next upgrade, A+~\cite{Aplus} will likely be operational for the O5
run in a few years~\cite{Scenarios}.
Virgo also has an upgrade planned for O5~\cite{Scenarios}.
A later upgrade of the existing LIGO facilities known as A$^\sharp$ is also
being planned~\cite{Asharp}, including a site in Aundha, India.
Going beyond existing facilities and into the next decade, plans are being made
to construct ``next generation'' detectors.
The main projects are the Einstein Telescope~\cite{Hild2011} in Europe and
Cosmic Explorer~\cite{Evans2021} in the United States.
Cosmic Explorer is planned to be like LIGO but larger, with L-shaped detectors
at two sites and either 20\,km or 40\,km arms.
The Einstein Telescope ET-D configuration is planned to be a triangle of
overlapping interferometers with 10\,km arms, resulting in a different response
to \ac{GW} polarizations.
A pair of 15\,km L-shaped detectors is also under consideration.
Its response to continuous wave signals is almost the same~\cite{ETScience}, so
here we consider the ET-D configuration only.
Like \citet{Hild2011} we consider linear polarization and convert it to the
equivalent response of an L-shaped detector.
Since we are comparing to existing upper limits marginalized over polarizations,
this is more appropriate than the circular polarization used in
\citet{Evans2021}.
We plot noise curves (strain amplitude spectral densities or asds) for these
detectors, and a recent LIGO noise curve~\cite{Capote2025} for reference, in
Fig.~\ref{f:noise}.

We consider the networks of these detectors listed in Table~\ref{t:networks}.
Most of these are a subset of the networks considered by \citet{Evans2023}.
Generally, H, L, and A stand for LIGO locations in Hanford, Livingston, and
Aundha (India) respectively, and are mostly taken to have A$^\sharp$
sensitivity.
The exception, the HLV network, should operate in O5 and means two LIGO
locations at A+ sensitivity plus Virgo at O5 sensitivity.
For continuous \acp{GW}, site locations generally have little effect since the
Earth's rotation samples different source-interferometer
alignments~\cite{Jaranowski1998}, and sensitivity is mainly a matter of noise
curves and integration times.
The 20LET, 40LET, and 40LA networks are omitted because they are similar to or
intermediate between 20LA and 4020ET, and plotting them makes the figures
visually crowded.

\begin{table}
\begin{tabular}{ll}
\hline
\hline
Network & Composition
\\
\hline
HLV & two A+ and Virgo (O5)
\\
HLA & three A$^\sharp$ (O6)
\\
20LA & CE 20\,km and two A$^\sharp$
\\
40LA & CE 40\,km and two A$^\sharp$
\\
4020A & CE 40\,km and CE 20\,km and A$^\sharp$
\\
4020ET & CE 40\,km and CE 20\,km and ET-D
\\
\hline
\hline
\end{tabular}
\caption{
\label{t:networks}
Acronyms for various detector networks.
They are as given in \protect\citet{Evans2023} except for HLV.
}
\end{table}

As we shall see, and as was seen in \citet{Evans2023} and \citet{Gupta2024},
building or upgrading detectors is generally preferable to operating existing
detectors for a longer time.
This can be understood in terms of simple scalings.
It is convenient to think in terms of detector arm length $L,$ with non-facility
improvements such as A+ being equivalent to increases in an effective $L.$
For a given signal, the observable \ac{GW} amplitude scales as $1/L$ and the
detectable range scales as $L.$
For compact binary coalescences, the only \ac{GW} signals observed to date, the
number of events is proportional to $L^3T,$ where $T$ is the observation
time~\cite{O4aCatalog}.
Therefore improving $L$ twofold is like improving $T$ eightfold.
For the stochastic background of \acp{GW}, which may be the next signal detected
by ground-based interferometers~\cite{Stochastic}, the signal-to-noise ratio
goes as $LT^{1/2}.$
Therefore improving $L$ twofold is like improving $T$ fourfold.
For continuous \acp{GW}, the signal-to-noise ratio ranges from $LT^{1/2}$ to
$LT^{1/4}$ depending on the search method~\cite{Wette2023}.
Therefore improving $L$ twofold can be like improving $T$ more than an order of
magnitude.

This paper is organized as follows:
In Sec.~\ref{s:cw} we briefly review the basics of continuous \ac{GW} emission
and data analysis.
We make estimates for the two broad classes of multi-messenger continuous
\ac{GW} sources, searches for known pulsars targeted on an \ac{EM} timing
solution and searches directed toward non-pulsing neutron stars, in
Secs.~\ref{s:pulsars} and~\ref{s:directed} respectively.
In Sec.~\ref{s:conclusion} we summarize our conclusions and give pointers toward
future work.

Throughout we use geometrized units, such that the speed of light and Newton's
gravitational constant are unity.

\section{Continuous gravitational waves}
\label{s:cw}

The first printed mention of continuous \acp{GW} from deformed neutron stars
appears to be \citet{Shklovskii1969}, with several others later that year as
summarized in \citet{Press1972}.
The topic is now well developed with many reviews, including a theoretical view
of various neutron star \acp{GW}~\cite{Glampedakis2018}, a theoretical and
observational overview~\cite{Jones2025}, a detailed study of
sensitivity~\cite{Wette2023}, a detailed summary of
observations~\cite{Riles2023}, and a brief but accessible survey~\cite{OwenRMP}.
These reviews include descriptions of the surprising variety of things that can
be learned even from a single detection, since that detection will have more
cycles than all the compact binary coalescences observed to
date~\cite{Jaranowski1999, O4aCatalog}.

Emission of \acp{GW} requires a changing quadrupole (or higher)
moment~\cite{Thorne1980}, and there are three main mechanisms for achieving this
with continuous waves from neutron stars: elastically or magnetically supported
mass quadrupoles~\cite{Shklovskii1969} (analogous to mountains) and current
quadrupoles~\cite{Lindblom1998} from long-lived $r$-mode
oscillations~\cite{Papaloizou1978a}.
We omit less likely emission mechanisms such as free precession~\cite{Jones2001}
and unstable $f$-modes~\cite{Glampedakis2018}, and speculative mechanisms such
as axion clouds around black holes summarized in \citet{Piccinni2022} and
\citet{Miller2026}.
We do, however, use a broad definition of ``neutron star,'' allowing for the
possibility that these stars may contain quark matter, meson condensates, or
other possibilities compatible with the uncertainties of low energy quantum
chromodynamics.
Neutron stars accreting matter from a binary companion are of special interest
as continuous \ac{GW} sources since accretion is a natural way of generating
asymmetries through electron capture layers and lateral temperature
gradients~\cite{Bildsten1998}, magnetic bottling of accreted
material~\cite{Melatos2005}, or the $r$-mode \ac{GW}
instability~\cite{Bildsten1998, Andersson1999b}.

Although the multipoles of a \ac{GW} source determine its
emission~\cite{Thorne1980}, continuous wave perturbations are typically quoted
in terms of other quantities.
The mass quadrupole is usually written as a fiducial ellipticity, given in terms
of the moment of inertia tensor $I$ by
\begin{equation}
\epsilon = \left( I_{xx} - I_{yy} \right) / I_{zz}.
\end{equation}
$R$-mode amplitudes are usually written in terms of a sort of fractional
velocity perturbation of the fluid, related to the Eulerian velocity
perturbation by~\cite{Lindblom1998}
\begin{equation}
\delta \vec{v} = \frac{\alpha} {\sqrt{6}} \Omega R (r/R)^2 r\vec{\nabla} \times
\left( r\vec{\nabla} Y_{22} \right) e^{i\omega t}
\end{equation}
in terms of the stellar angular spin frequency $\Omega,$ radius $R,$ and angular
mode frequency $\omega,$ where $Y_{22}$ is the standard spherical harmonic.
The common quantities $\epsilon$ and $\alpha$ are related to the proper mass
quadrupole and current quadrupole~\cite{Thorne1980} by~\cite{Owen2010}
\begin{equation}
\left| I^{22} \right| = \sqrt{ \frac{8\pi} {5} } I_{zz} \epsilon,
\qquad
\left| S^{22} \right| = - \frac{ 32\sqrt{2}\pi } {15} \alpha M\Omega R^3
\tilde{J},
\end{equation}
where $M$ is the star's mass.
Good mass measurements range from about 1.2\,$M_\odot$ to a little more than
2\,$M_\odot,$ while radius measurements are roughly
11--14\,km~\cite{Lattimer2021}.
The quantity $\tilde J \simeq 0.0164$ varies little with the mass and equation
of state.

Continuous \ac{GW} data analysis is based on matched filtering or near
equivalents, much like data analysis of compact binary coalescences.
A thorough explanation of the ``$\mathcal{F}$-statistic'' variant is given in
\citet{Jaranowski1998}.
The main difference is that coherent integration times are much longer, even in
computationally limited searches that semi-coherently combine relatively short
stretches of data.
Coherent integration times of most searches are at least a few days, even in the
initial stages of all sky surveys, and can run years in known pulsar searches
such as Ref.~\cite{S5Crab}.
Each day the detectors rotate with the Earth and change orientations, modulating
the amplitude and phase of each detector's response and inducing a small
time-dependent Doppler shift depending on the sky position.
The yearly motion of the detectors with the earth's orbit induces a larger
time-dependent Doppler shift.
Thus continuous waves can achieve tremendous sky location accuracy and sample
different polarizations even with a single L-shaped detector.
Corrected for these effects, the basic signal model is a slow frequency drift
approximated by a power series in time.
\ac{EM} observations of pulsars indicate that there can be additional stochastic
variation known as timing noise, and young pulsars occasionally have abrupt
frequency jumps known as glitches.
Thus \ac{GW} searches for known pulsars use frequent \ac{EM} timing.
Since timing noise (comparable to the noisiest known pulsars) becomes an issue
only for integration times of years~\cite{Ashton2015}, searches for non-pulsing
neutron stars are not greatly affected by it.
Unseen glitches may be more of a concern for these searches if they use
coherence times of months~\cite{Ashton2017}.
Similar fluctuations might arise for saturated $r$-modes~\cite{Owen2010}.

Continuous wave amplitudes are typically expressed in terms of the
\textit{intrinsic strain} $h_0$~\cite{Jaranowski1998}.
Amplitude modulations due to detector motion make the usual practice of quoting
a detector response (strain measured by a given detector) problematic.
Instead $h_0$ is reconstructed in terms of the amplitude of the oscillating
metric perturbation tensor.
(This is generally assumed to be constant, but if it varies searches are
sensitive to the root mean square value.)
It has a simple relation to the \ac{GW} luminosity $dE/dt$ via
\begin{equation}
\frac{dE}{dt} = \frac{1}{10} \omega^2 D^2 h_0^2,
\end{equation}
where $D$ is the distance to the source and now $\omega$ is the \ac{GW} angular
frequency regardless of emission mechanism.

Intrinsic strain can also be related simply to the amplitude of the asymmetry
generating the gravitational waves.
For a non-axisymmetric ellipticity the relation to the source perturbation
is~\cite{Jaranowski1998}
\begin{equation}
\label{h0eps}
h_0  = \frac{\omega^2 I_{zz}} {D} \epsilon = 4.23\times10^{-24} \left(
\frac{\mathrm{1\,kpc}} {D} \right) \left( \frac{\mathrm{1\,s}} {P} \right)^2
\epsilon,
\end{equation}
where $P$ is the stellar spin period, the spin frequency is $1/P,$ and we assume
\ac{GW} emission at twice the spin frequency.
The fiducial form on the rhs of Eq.~(\ref{h0eps}) assumes a moment of inertia
$I_{zz} = 10^{45}$\,g\,cm$^2,$ which is rather conservative since it could be
more than 4 times higher~\cite{O3J0537Rmodes}.
The numerical factor is uncommon in the literature but is easily checked
against, say, Sec.~3 of a well known result~\cite{S5Crab}.
The analogous relation for an $r$-mode is~\cite{Owen2010}
\begin{equation}
\label{h0alpha}
h_0 = \sqrt{ \frac{8\pi}{5} } \frac{\omega^3 MR^3 \tilde{J}} {D} \alpha =
1.23\times10^{-28} \left( \frac{\mathrm{1\,kpc}}{D} \right) \left(
\frac{\mathrm{1\,s}}{P} \right)^3 \alpha.
\end{equation}
The fiducial version at right assumes $M = 1.4\,M_\odot$ and $R \simeq
11.7$\,km, consistent with the fiducial moment of inertia for a Newtonian
polytrope.
The gravitational wave frequency is $A$ times the spin frequency, where $A$
depends on the neutron star equation of state and mass, and is predicted to lie
within the range 1.39---1.64~\cite{Idrisy2015, Ghosh2023}.
We use $A=1.5$ as the fiducial value since it is about the middle of that range.
The fiducial $h_0$ is proportional to $A^3.$

We express sensitivities of future searches in terms of upper limits achievable
at a certain (high) confidence level or (low) false dismissal probability.
This is standard practice in the literature, and it allows direct comparison
with previous searches, all of which have produced upper limits rather than
detections.
However, it is a conservative estimate.
Often it is marginalized over inclination and polarization angles, meaning that
for instance 50\% detection probability occurs at a significantly lower strain.
This marginalization also means that high confidence upper limits are dominated
by linearly or almost linearly polarized signals, which is why we use linear
polarization for the equivalent ET-D noise curve.

In recent years, continuous \ac{GW} searches are often characterized by
\textit{sensitivity depth} $\mathcal{D},$ defined by~\cite{Behnke2015,
Dreissigacker2018}
\begin{equation}
\label{depth}
h_0^\mathrm{UL} = \sqrt{S_h} / \mathcal{D},
\end{equation}
where $\sqrt{S_h}$ is the harmonic mean of the noise asd over different
detectors and times.
This $\mathcal{D}$ is useful for extrapolating to future detectors because it
includes every factor but the noise spectrum.
Its disadvantage is that it hides the dependence on data live time $T.$
It also depends on other practical choices of the data analysis method and code.
Roughly speaking, for a single coherent integration $\mathcal{D}$ scales as
$T^{1/2}$ and for semi-coherent searches it scales as $T^{1/4}.$

\section{Known pulsars}
\label{s:pulsars}

Known pulsar searches use the neutron star's sky position, frequency, frequency
evolution, and orbital parameters (if any) derived from \ac{EM} observations.
These searches may precisely track a known pulsar timing solution (such as
searching multiples of the evolving spin frequency inferred from radio data) or
search a narrow band of frequencies and frequency derivatives around a timing
solution---see \citet{O4aPulsars} for a recent example of both.
Narrow band searches may be done if the pulsar timing is not precisely known at
the desired epoch or if the search is to be sensitive to emission mechanisms
more complicated than a simple ``mountain.''
For example, the crust and core of the neutron star might spin at slightly
different rates, jumping into sync at glitches~\cite{S5Crab}.
Somewhat wider narrow band searches are needed to target $r$-modes, whose
frequencies are related to the spin frequency but depend on the neutron star
equation of state~\cite{Caride2019}.

Known pulsar searches are the most sensitive continuous \ac{GW} searches of a
given data set for two reasons.
First, being limited to a single point or a small region in parameter space,
they can coherently integrate all available data for modest computational cost.
In some cases they have even been run as quick follow-ups to electromagnetic
observations of newly discovered pulsars~\cite{Clark2023, Thongmeearkom2026}.
Second, since they make one or relatively few statistically independent trials
of the noisy data, their thresholds for statistically significant detection can
be relatively low.

The first sensitivity milestone a known pulsar search needs to reach is the
\textit{spin-down limit,} or the upper limit set by the (unrealistic) assumption
that all of the \ac{EM} observed spin-down is produced by \ac{GW} emission.
The spin-down limit on intrinsic strain, assuming that the \acp{GW} are emitted
by a mass quadrupole at twice the spin frequency, is~\cite{O4aPulsars}
\begin{equation}
\label{h0sd2}
h_0^\mathrm{sd} = \frac{1}{D} \sqrt{ \frac{5I_{zz}}{2}
\frac{\left|\dot{f}\right|} {f} }
= 8.06\times10^{-19} \left( \frac{\mathrm{1\,kpc}} {D} \right) \sqrt{ \dot{P}
\frac{\mathrm{1\,s}} {P} }.
\end{equation}
Here $f$ is the \ac{GW} frequency, assumed to be twice the spin frequency, and
$\dot{P}$ is the (dimensionless) spin period derivative.
Via Eq.~(\ref{h0eps}), $h_0^\mathrm{sd}$ corresponds to an ellipticity upper
limit of
\begin{equation}
\epsilon_\mathrm{sd} = \sqrt{ \frac{5} {32\pi^4 I_{zz}}
\frac{\left|\dot{f}\right|} {f^5} }
= 1.91\times10^5 \sqrt{ \dot{P} \left( \frac{P} {\mbox{1\,s}}
\right)^3 }.
\end{equation}

For $r$-mode emission there is a slightly different spin-down limit on intrinsic
strain~\cite{Owen2010},
\begin{equation}
\label{h0sdr}
h_0^\mathrm{sd} = \frac{1}{D} \sqrt{ \frac{10I_{zz}} {A^2} \frac{\left|
\dot{f} \right|} {f} }
= 1.07\times10^{-18} \left( \frac{\mathrm{1\,kpc}}{D} \right) \sqrt{
\dot{P} \frac{\mathrm{1\,s}} {P} } .
\end{equation}
In the first equation we have generalized to arbitrary $A$ from Eq.~(26) of
\citet{Owen2010}, which uses the original Newtonian slow rotation approximation
$A=4/3$~\cite{Papaloizou1978a}, and in the second we have again used a fiducial
$A=1.5.$
The corresponding spin-down limit on $r$-mode amplitude $\alpha$ from
Eq.~(\ref{h0alpha}) is
\begin{eqnarray}
\alpha_\mathrm{sd} &=& \frac{1}{A} \frac{5} {16\pi^{7/2}} \sqrt{ \frac{I_{zz}
\left| \dot{f} \right|} {f^7} } \frac{1} {\tilde{J} MR^3}
\nonumber\\
&=& \frac{1}{A^4} \frac{5} {81\pi^{7/2}} \sqrt{ I_{zz} \dot{P} P^5} \frac{1}
{\tilde{J} MR^3}
\nonumber\\
&=& 8.80\times10^9 \sqrt{ \dot{P} \left( \frac{P} {\mathrm{1\,s}} \right)^5}
\end{eqnarray}
Equation~(27) of Ref.~\cite{Owen2010} is the same but uses $A=4/3,$ while here
we use a fiducial value of $A=3/2.$
Note that the fiducial form of our equation has a hidden $1/A^4$ scaling when
given in terms of $P$ rather than $f.$

Much of the literature seems optimistic because it quotes spin-down limits for
detectability.
The first continuous \ac{GW} search to beat the spin-down limit on a known
pulsar was a long time ago~\cite{S5Crab}, and dozens have been surpassed since
then~\cite{O4aPulsars}.
Hundreds of spin-down limits will be accessible to next generation
detectors~\cite{Branchesi2023, Riles2023}.

But the spin-down limit on \ac{GW} amplitude is just a first milestone of
sensitivity.
For most pulsars the spin-down limit on $h_0$ corresponds to an unfeasibly large
ellipticity, unless we consider extreme theories such as quark
matter~\cite{Owen2005, S5Crab} or magnetar-strength internal magnetic
fields~\cite{S5Crab}.
And while a neutron star \textit{could} have that ellipticity, in most cases we
do not know why it \textit{should.}
Neutron stars are stably stratified, and thus cannot support plate tectonics and
terrestrial mountain building processes~\cite{Glampedakis2018}.
Magnetar-strength magnetic fields are likely very rare for stars still spinning
at millisecond periods, since such a field spins the star down quickly.
For more conservative and realistic detectability estimates, we turn first to a
case where a neutron star \textit{should} emit at a certain amplitude.

\subsection{Magnetically deformed millisecond pulsars}

\begin{figure}
\includegraphics[width=0.45\textwidth]{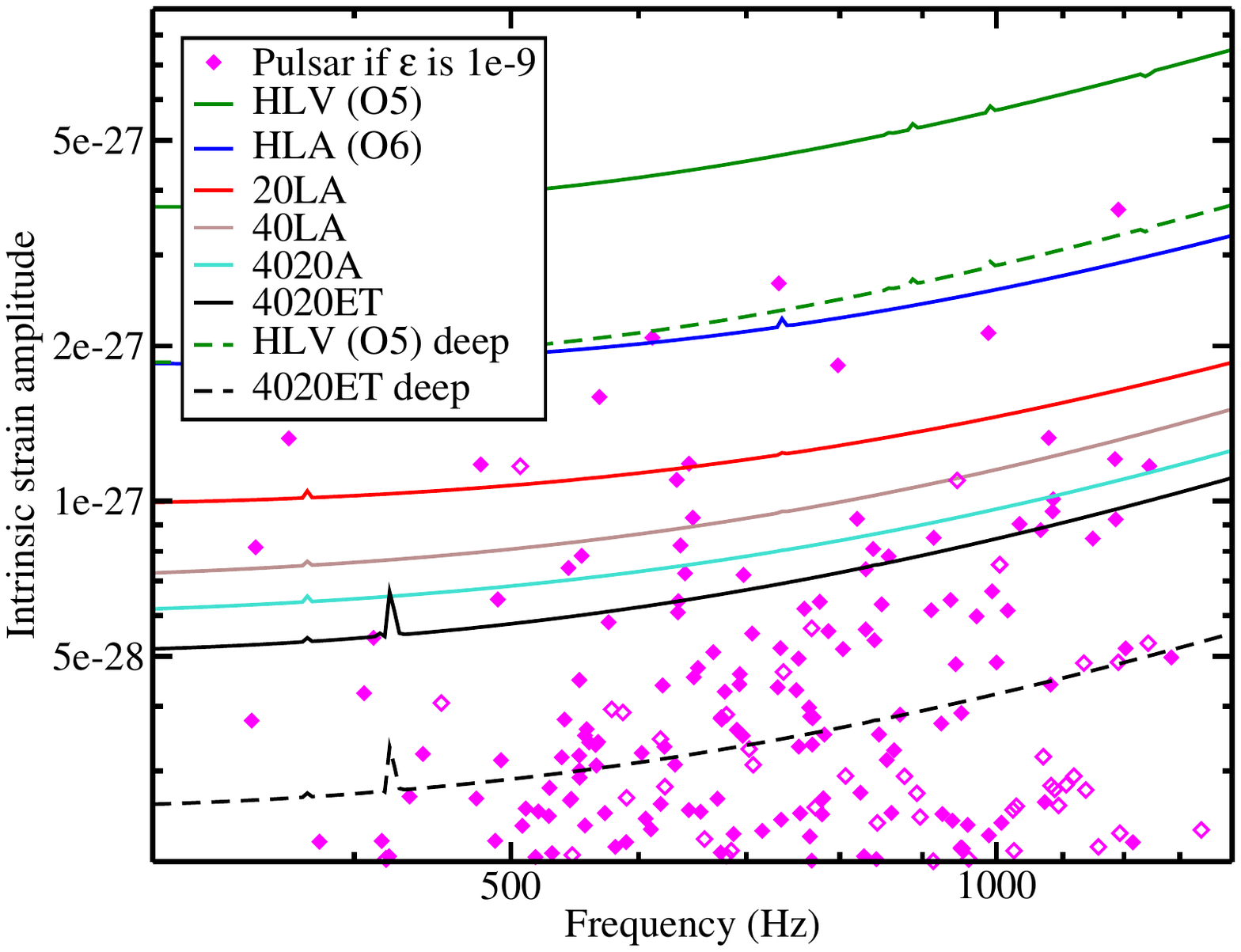}
\caption{
\label{f:eps1e-9}
Millisecond pulsars that \textit{should} be detected by various networks.
Pink diamonds show intrinsic strain for known pulsars assuming an ellipticity of
$10^{-9}$, except for rare cases where the spin-down limit is slightly lower and
is shown instead.
Filled diamonds indicate pulsars for which the spin-down is known.
Hollow diamonds indicate pulsars for which it is not known, and therefore the
strain plotted may be too high.
Solid curves show sensitivities for various interferometer networks assuming a
sensitivity depth of 500\,Hz$^{-1/2}.$
The two dashed curves assume a sensitivity depth of 1000\,Hz$^{-1/2},$ which is
achievable with long, stable runs and good pulsar timing.
}
\end{figure}

\citet{Woan2018} argue that millisecond pulsars \textit{should} have a minimum
ellipticity, an argument with both observational and theoretical justifications.
Pulsars are commonly plotted on the $P$--$\dot{P}$ diagram (see for example
Fig.~1 of \citet{Woan2018}) to distinguish various populations and estimate
(external dipole) magnetic fields and other properties.
Millisecond pulsars, so called because their spin periods may be as short as
milliseconds, are a distinct population at small $P$ and small $\dot{P}.$
Inspection of the millisecond pulsar population in the diagram suggests a
boundary in the $P$--$\dot{P}$ plane corresponding to a constant quadrupole
($\dot{P} \propto P^{-5}$) rather than a constant dipole ($\dot{P} \propto
P^{-7}$) as for other pulsars, and statistical analysis shows strong evidence
for it~\cite{Woan2018}.
Such a boundary is hard to explain with exterior magnetic fields, which appear
to be dominated in millisecond pulsars by dipoles of order
$10^8$--$10^9$\,G~\cite{ATNF}.
But it has a natural explanation if an internal dipole magnetic field of order
$10^{11}$\,G exists, producing a mass quadrupole ellipticity of order $10^{-9}$.
Such an internal field is consistent with the exterior dipoles observed in the
spin-downs of young neutron stars~\cite{ATNF}.
It is also consistent with the standard theory of millisecond pulsar formation,
that a long period of accretion buries the initial magnetic field, leaving only
a small remnant as the external dipole observed today~\cite{Alpar1982}.
The observation of \ac{EM} pulses indicates that the field axis is misaligned
with the spin axis, leading to a misaligned mass quadrupole and \ac{GW}
emission.
(Even beyond the biased sample of pulsars, some misalignment is expected to be
common, and some young pulsars are driven to orthogonal alignments---see for
example \citet{DallOsso2017}, \citet{Lander2018}, and references therein.)

There are some caveats.
The observed cutoff at $\epsilon\sim10^{-9}$ is not sharp, especially with
more recent data from the \ac{ATNF} catalog~\cite{ATNF}.
For a given field strength the ellipticity can vary by a factor of a few
depending on its configuration.
Some stars may be born with weaker internal fields.
Internal fields may decay on timescales of millions of years~\cite{Pons2019}
rather than billions~\cite{Goldreich1992}.
Maybe there is an unknown selection effect.
But $\epsilon\sim10^{-9}$ remains interesting because it starts to tell us when
even nondetection could shed some light on these issues.

In Fig.~\ref{f:eps1e-9} we plot intrinsic strains of known pulsars assuming
$\epsilon=10^{-9}$ (pink diamonds), compared to sensitivities of fully targeted
searches with various detector networks (curved lines).
All detectable pulsars for this ellipticity are millisecond pulsars, since the
scaling of Eq.~(\ref{h0eps}) favors high \ac{GW} frequencies.
Solid detectability curves are obtained from Eq.~(\ref{depth}) assuming a
sensitivity depth of 500\,Hz$^{-1/2},$ which is a typical value for fully
targeted searches~\cite{Wette2023} including the most recent~\cite{O4aPulsars}.
Dashed curves show the most optimistic scenario, assuming a sensitivity depth of
1000\,Hz$^{-1/2},$ which is about the best on record~\cite{Wette2023} and has
been achieved with long, stable runs and plenty of \ac{EM} timing data.
To avoid visual crowding, we plot these only for the earliest and the best case
networks.
We assume \ac{GW} emission at twice the spin frequency with amplitude given by
Eq.~(\ref{h0eps}).
Keep in mind that the minimum ellipticity is a rough number and can easily vary
by a factor two or so from pulsar to pulsar due to differing masses, field
configurations, \textit{etc.}
Also the $h_0$ for a given ellipticity depends on distance, which sometimes
yields a different result when estimated by different methods.
Pulsars and their parameters are taken from \ac{ATNF} catalog \cite{ATNF}
version 2.7.0 processed with the \texttt{psrqpy} Python package~\cite{psrqpy}.
In a few cases the spin-down limit is slightly worse than Eq.~(\ref{h0eps}), so
we have plotted $h_0^\mathrm{sd}$ for those pulsars instead.
The hollow diamonds indicate pulsars for which there is no $h_0^\mathrm{sd}$
since $\dot{P}$ is not known.
We must keep in mind that for these pulsars Eq.~(\ref{h0eps}) may be an
overestimate, although that should be rare based on the cases where $\dot{P}$ is
known.
In cases where the \ac{ATNF} catalog lists an intrinsic $\dot{P}$ value (with
kinematic effects corrected), we have used that instead of the raw $\dot{P}.$

Taking the numbers at face value, Fig.~\ref{f:eps1e-9} shows that---if the
minimum ellipticity argument is correct---three pulsars should be detectable
already in O6 or even (with a long stable run) in O5.
With even the smallest version of Cosmic Explorer, the number jumps to ten for a
conservative sensitivity depth.
(Although not plotted, the number is similar with the Einstein Telescope as the
only next generation detector.
It is also similar for HLA (O6) with an optimistic sensitivity depth.)
With the best next generation network (4020ET), the number of detectable
millisecond pulsars is in the mid-30s even for a conservative sensitivity depth,
and roughly double that for an optimistic sensitivity depth.
These numbers are slightly different from the estimates in \citet{Evans2023,
Gupta2024}.
Partly this is due to the assumed sensitivity depths and partly it is due to an
updated pulsar catalog with more good pulsars discovered even in such a short
time.

In the era of Einstein Telescope and Cosmic Explorer there should be several new
pulsars for each pulsar known now, and that factor is likely to be even higher
for millisecond pulsars~\cite{Smits2009}.
Therefore the number detected in an optimistic scenario could reach into the
hundreds.
At some point, well before then, even a lack of detections becomes very
interesting.
If the ``best'' pulsar in Fig.~\ref{f:eps1e-9} is not detected soon, one can
always argue that it is just one of the pulsars born with a weak field, or is on
the bad end of the range of moment of inertia or one of the other uncertain
factors of order unity.
But if many pulsars become detectable at $\epsilon = 10^{-9}$ and are not
detected, it calls into question the standard theory of millisecond pulsar
formation, and that would be a very interesting result in itself.

\begin{table}
\begin{tabular}{lcrcr}
\hline
\hline
\multicolumn{1}{c}{JName} &
\multicolumn{1}{c}{$h_0$} &
\multicolumn{1}{c}{$f$} &
\multicolumn{1}{c}{$-\dot{f}$} &
\multicolumn{1}{c}{$D$}
\\
&
\multicolumn{1}{c}{if $\epsilon=10^{-9}$} &
\multicolumn{1}{c}{(Hz)} &
\multicolumn{1}{c}{$\left( \mathrm{Hz}\, \mathrm{s}^{-1} \right)$} &
\multicolumn{1}{c}{(pc)}
\\
\hline
J1036\textminus4353 & $3.67\times10^{-27}$ & 1190 & $4.52\times10^{-15}$ & 408
\\
J0605+3757 & $2.64\times10^{-27}$ & 733 & $1.27\times10^{-15}$ & 215
\\
J1720\textminus0533 & $2.07\times10^{-27}$ & 612 & $1.53\times10^{-15}$ & 191
\\
J0955\textminus3947 & $2.12\times10^{-27}$ & 989 & $1.85\times10^{-14}$ & 488
\\
J0646\textminus5455 & $1.83\times10^{-27}$ & 798 & $5.77\times10^{-16}$ & 367
\\
J1709\textminus0333 & $1.59\times10^{-27}$ & 568 & $3.02\times10^{-16}$ & 214
\\
J0711\textminus6830 & $1.32\times10^{-27}$ & 364 & $9.89\times10^{-16}$ & 106
\\
J1737\textminus0811 & $1.18\times10^{-27}$ & 479 & $9.10\times10^{-16}$ & 206
\\
J0636\textminus3044 & $1.17\times10^{-27}$ & 507 & --- & 233
\\
J1012\textminus4235 & $1.18\times10^{-27}$ & 645 & $1.36\times10^{-15}$ & 372
\\
\hline
\hline
\end{tabular}
\caption{
\label{tab:msps}
Top ten millisecond pulsars from Fig.~\protect\ref{f:eps1e-9} in order of likely
detection, given the assumptions and caveats mentioned in the text.
Here $\epsilon$ is assumed to be $10^{-9}$ or the spin-down limit, whichever is
lower.
The \ac{GW} frequency $f$ is assumed to be twice the spin frequency.
Its time derivative $\dot{f}$ is the intrinsic one when that is available.
These and the distance $D$ are from v2.7.0 of the \ac{ATNF}
catalog~\protect\cite{ATNF}.
}
\end{table}

In Table~\ref{tab:msps} we list the ten known millisecond pulsars likely to be
detected first, in order of likely detection (taking the numbers in
Fig.~\ref{f:eps1e-9} at face value).
References to details of these pulsars are linked from the \ac{ATNF}
catalog~\cite{ATNF}.
Due to the scalings of Eq.~(\ref{h0eps}), this list consists of fast nearby
pulsars, and hence has much overlap with the list considered for long duration
post-glitch \ac{GW} searches by~\citet{Yim2024}.
Most of these pulsars are in binaries, and many of them are ``redback'' or
``black widow'' pulsars which are significantly interacting with their
companions.
On the one hand these ``spider pulsars'' tend to need frequent timing since
their spins and/or orbits often show fluctuating torques.
On the other hand an interplay of thermal and magnetic effects may make them
good \ac{GW} emitters~\cite{Hutchins2023}.
Most of these pulsars are seen in gamma rays, and many were found in recent
surveys such as \citet{Clark2023} following up gamma-ray point sources.
This indicates that new discoveries are continuing to come in at a high rate.
The current best pulsar was not discovered at the time of \citet{Woan2018}.
Most of the pulsars listed have not been searched at all for gravitational
waves.
The three exceptions, such as J0605+3757 which was searched starting in the
``initial detector'' era~\cite{S6Pulsars}, were not searched in the first
O4 paper~\cite{O4aPulsars}.

Table~\ref{tab:msps} will need updating many times before next generation
\ac{GW} and radio detectors are built, and \ac{GW} observatories will need to
maintain and strengthen their links with \ac{EM} observatories.

\subsection{Young neutron stars}

\begin{figure}
\includegraphics[width=0.45\textwidth]{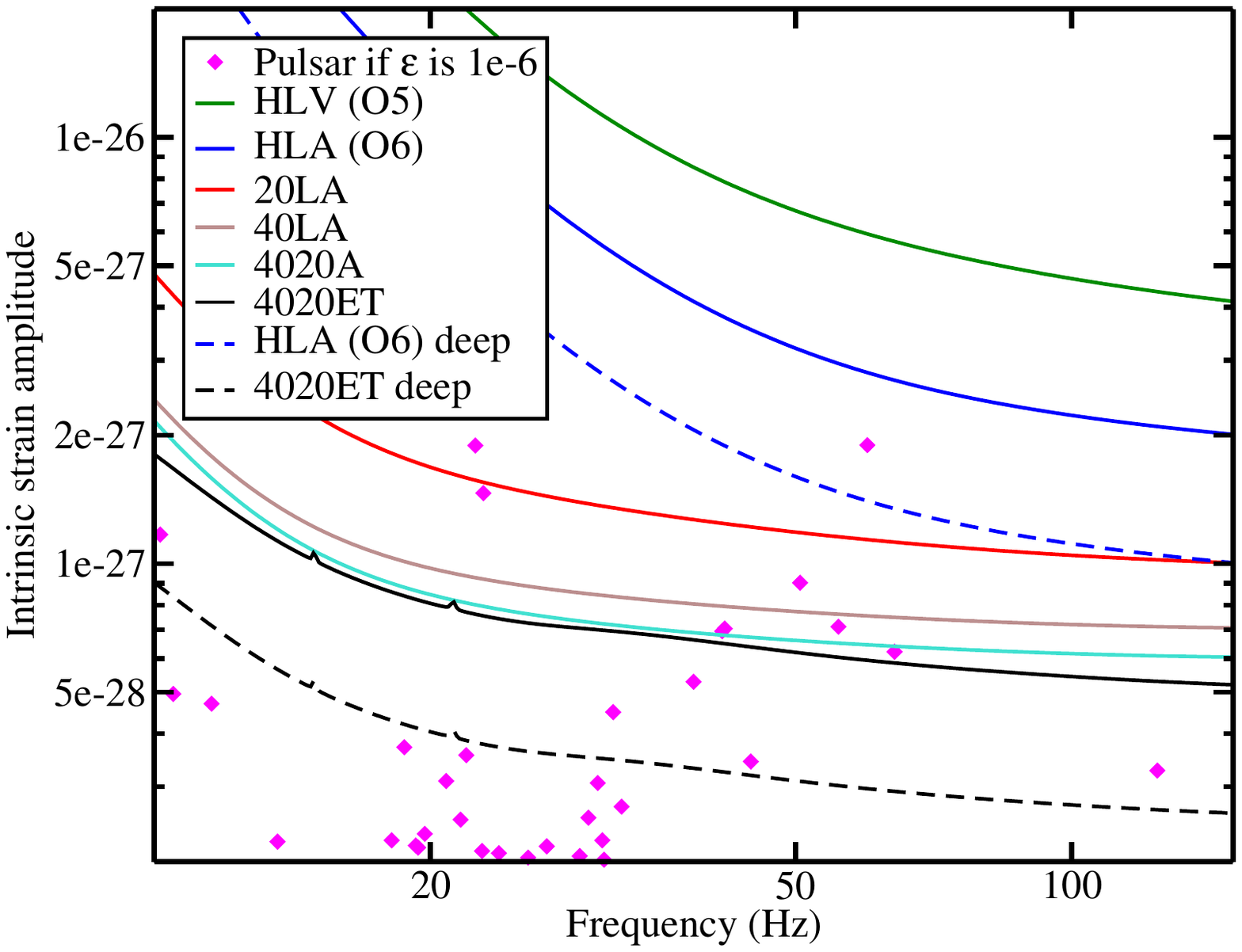}
\caption{
\label{f:eps1e-6}
Young pulsars that \textit{could} be detected by various networks.
Pink diamonds are as in the previous Figure, but assuming an ellipticity of
$10^{-6}.$
Here we plot only pulsars whose spin-downs are known and consistent with that
ellipticity and whose characteristic ages are less than $10^7$\,yr.
Solid and dashed curves have sensitivity depths as in the previous Figure.
}
\end{figure}

We now turn to a class of objects which \textit{could} emit at detectable
levels---young pulsars with rapid spin-downs that do not rule out a high
ellipticity.
Unlike for old millisecond pulsars, for young pulsars we do not have a generic
argument for why they \textit{should} commonly have a certain ellipticity.
They \textit{could} be highly deformed by elastic stresses (perhaps via the
asymmetry of the supernova that gave them birth) or magnetic stresses (if their
internal fields are very high).
If young pulsars have ellipticities at the minimum argued for millisecond
pulsars~\cite{Woan2018}, their \acp{GW} are undetectable.
But if they have much higher ellipticities, the picture becomes much more
interesting.

Estimates of maximum elastic deformations have varied over the years, but for
normal neutron stars (as opposed to quark matter and other exotica) they tend to
be on the order of $10^{-5}$--$10^{-6}$~\cite[e.g]{Morales2022}.
For quark matter the estimates can go several orders of magnitude
higher~\cite{Owen2005}.
That indicates that some pulsars, starting with the Crab, \textit{could} have
been detected by previous searches.
But because there is no reason they \textit{should} be near their maximum
ellipticity, their non-detection up to now does not rule out anything universal
such as the existence of exotic high density phases of matter.
Any star may simply be strained far below its maximum.

Magnetic deformations are more interesting in this context.
The internal field required to generate $\epsilon$ of order $10^{-6}$ is of
order $10^{14}$--$10^{15}$\,G~\cite{Lander2013}, much higher than the external
dipole fields inferred from observed spin-downs~\cite{ATNF}.
But there are arguments for creating such fields, and there is a more
observationally founded argument.
\citet{DallOsso2017} argue that long-term timing observations of the Crab
pulsar, which are usually interpreted to mean that its magnetic axis is slowly
moving away from the spin axis~\cite{Lyne2013}, could indicate an internal
magnetic field resulting in an ellipticity of a few times $10^{-6}.$
There are some caveats for the model~\cite{Lander2018}, but it is another
indication that $\epsilon=10^{-6}$ is an interesting number.

In Fig.~\ref{f:eps1e-6} we plot young pulsars that \textit{could} be detectable
if they have $\epsilon = 10^{-6}.$
The sensitivity curves are the same as Fig.~\ref{f:eps1e-9} for the same
reasons, except that here we have plotted a deep curve for HLA (O6) rather than
HLV (O5) because it makes more difference for detectability.
The pulsars plotted are different:
The values of $h_0$ shown are for $\epsilon = 10^{-6}$ if that is compatible
with the spin-down, and zero otherwise (no empty diamonds).
We plot only pulsars with characteristic ages $P/(2\dot{P}) < 10^7$\,yr (the
plot is not highly sensitive to the precise cutoff).
A plot without the age cutoff would include many spider pulsars,
whose $\dot{P}$ values tend to be heavily contaminated by other torques such as
the propeller effect~\cite[e.g]{Misra2025}.

The message of Fig.~\ref{f:eps1e-6} is that two pulsars (the Crab and Vela)
could be detected, without invoking exotic composition or extreme internal
magnetic fields, with even one next generation detector and a conservative
sensitivity.
With an optimistic sensitivity depth the Crab is accessible as early as O6.
Several more young pulsars are accessible with better detectors and/or deeper
searches.
A list of the top few is given in Table~\ref{tab:young}.
It is a group of generally Crab-like isolated pulsars, with the exception of one
that orbits a Be companion.
All have been searched previously for gravitational waves, though a few were not
in O4 \cite{O4aPulsars}.
The rightmost pulsar plotted in Fig.~\ref{f:eps1e-6}, and not included in
Table~\ref{tab:young}, is J0537\textminus6910.
It glitches so frequently that continuous timing information is crucial, and is
also observed only in X-rays, so future missions such as \ac{AXIS} are
important.

\begin{table}
\begin{tabular}{lcrcr}
\hline
\hline
\multicolumn{1}{c}{JName} &
\multicolumn{1}{c}{$h_0$} &
\multicolumn{1}{c}{$f$} &
\multicolumn{1}{c}{$-\dot{f}$} &
\multicolumn{1}{c}{$D$}
\\
&
\multicolumn{1}{c}{if $\epsilon=10^{-6}$} &
\multicolumn{1}{c}{(Hz)} &
\multicolumn{1}{c}{$\left( \mathrm{Hz}\, \mathrm{s}^{-1} \right)$} &
\multicolumn{1}{c}{(pc)}
\\
\hline
J0534+2200 & $1.90\times10^{-27}$ & 59.9 & $7.55\times10^{-10}$ & 2000
\\
J0835\textminus4510 & $1.89\times10^{-27}$ & 22.4 & $3.13\times10^{-11}$ & 280
\\
J0940\textminus5428 & $1.46\times10^{-27}$ & 22.8 & $8.58\times10^{-12}$ & 377
\\
J1952+3252 & $9.02\times10^{-28}$ & 50.6 & $7.48\times10^{-12}$ & 3000
\\
J1913+1011 & $7.12\times10^{-28}$ & 55.7 & $5.24\times10^{-12}$ & 4609
\\
J1302\textminus6350 & $7.05\times10^{-28}$ & 41.9 & $2.00\times10^{-12}$	& 2632
\\
J1813\textminus1246 & $6.95\times10^{-28}$ & 41.6 & $1.52\times10^{-11}$ & 2635
\\
J1400\textminus6325 & $6.22\times10^{-28}$ & 64.1 & $8.00\times10^{-11}$	& 7000
\\
\hline
\hline
\end{tabular}
\caption{
\label{tab:young}
Like Table~\protect\ref{tab:msps} but for the top eight pulsars from
Fig.~\protect\ref{f:eps1e-6}.
The top two pulsars are commonly known as the Crab and Vela.
}
\end{table}

\subsection{R-modes}

\begin{figure}
\includegraphics[width=0.45\textwidth]{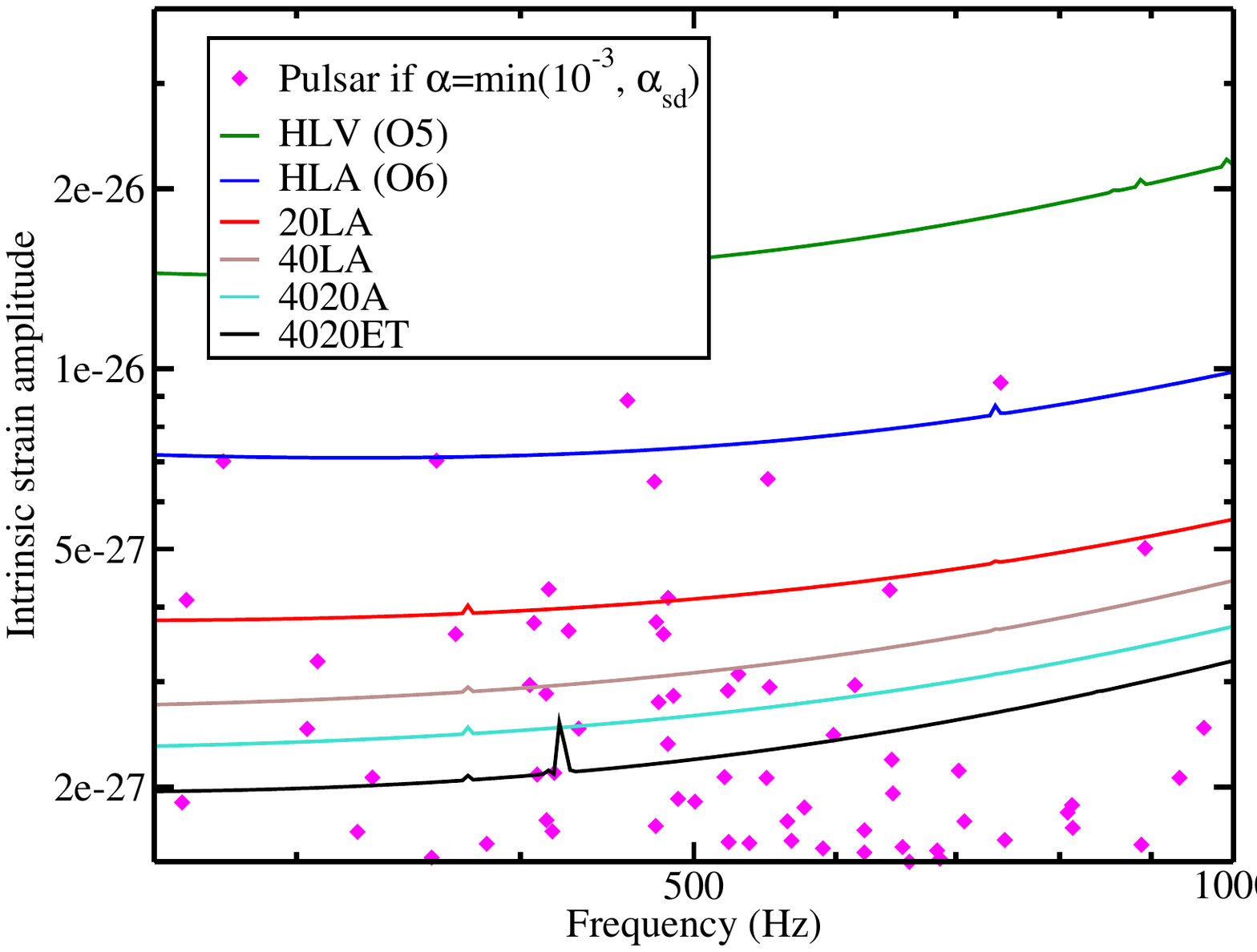}
\caption{
\label{f:rmodes}
Pulsars that could be detectable via $r$-mode \ac{GW} emission.
Pink diamonds are as the previous two figures, but assuming an $r$-mode with
\ac{GW} emission at 1.5 times the spin frequency and amplitude $\alpha$ of
$10^{-3}$ or the spin-down limit, whichever is lower.
Detectability curves assume a sensitivity depth of 100\,Hz$^{-1/2}.$
}
\end{figure}

The argument that $r$-modes \textit{should} emit \acp{GW} at a certain amplitude
boils down to a \ac{GW}-driven instability~\cite{Andersson1998, Friedman1998}
which survives the presence of basic neutron star viscosity~\cite{Lindblom1998,
Andersson1999a, Lindblom1999}.
That is, a small $r$-mode is driven rather than damped by \ac{GW} emission due
to the Chandrasekhar-Friedman-Schutz instability~\cite{Chandrasekhar1970,
Friedman1978}.
The instability probably survives many other complicated damping mechanisms
reviewed \textit{e.g.} by \citet{Glampedakis2018}, both in newborn neutron
stars~\cite{Owen1998} and in rapidly accreting neutron stars in
\acp{LMXB}~\cite{Bildsten1998, Andersson1999b}.
Eventually the $r$-mode amplitude saturates, probably due to nonlinear
hydrodynamics~\cite{Owen1998, Arras2003}, at a value related to viscous damping
of other modes and varying from scenario to scenario.
In terms of $\alpha$ the saturation amplitude is probably at most $10^{-3}$ in
young neutron stars \cite{Bondarescu2009, Owen2010} and much less in older
millisecond pulsars with more damping, maybe as low as $10^{-6}$
\cite{Bondarescu2007}.
The saturation amplitude is small enough that $r$-mode \ac{GW} emission could
last thousands of years in newborn neutron stars~\cite{Owen1998, Arras2003,
Bondarescu2009}, for the duration of rapid accretion in the \ac{LMXB} phase, and
perhaps even longer in millisecond pulsars~\cite{Reisenegger2003} after the
\ac{LMXB} phase.
Indeed all millisecond pulsars might have unstable $r$-modes at some
level~\cite{Bondarescu2013}.
In both young and old objects the star spins down and cools after birth or
accretion, reducing the driving and increasing the damping until the instability
ends and eventually the stabilized mode damps away.

While the $r$-mode instability looks intriguing from a theoretical point of
view, there are caveats.
Observations of rapidly accreting neutron stars suggest that most of them are in
the originally predicted instability region, yet they do not show signs of
continuous instability.
This has not changed since the beginning~\cite{Andersson1999b} and indicates
that some of the complicated physics of $r$-mode damping is poorly understood.
Accreting neutron stars may have $r$-modes only intermittently~\cite{Levin1999,
Spruit1999}.
In some post-accreting millisecond pulsars the temperature is too low to be
consistent with the heating $r$-modes would cause~\cite{Boztepe2020}.
Although $r$-modes have been suggested as explanations for the overall spins of
young and old neutron stars, observed braking indices (combinations of frequency
derivatives)~\cite{ATNF} suggest that $r$-mode torque is responsible for only a
small fraction of spin-down in most cases.
That and quasi-universal scaling arguments suggest that PSR~J0537\textminus6910
might be the only viable candidate among young neutron stars~\cite{Alford2014,
Andersson2018}.
Note that it does not appear in Fig.~\ref{f:rmodes} because its
$\alpha_\mathrm{sd}$ is much higher than the maximum theoretical $\alpha$ of
order $10^{-3}.$
Saturation amplitudes are a complicated topic and have really been only
addressed by one research group with one set of
techniques~\cite[e.g.]{Bondarescu2009}.
A fundamental issue is that the saturation amplitude is known only very roughly.
Therefore, in spite of the instability that \textit{should} drive $r$-modes
toward saturation, we consider these sources that \textit{could} be detectable.

In Fig.~\ref{f:rmodes} we plot pulsars that \textit{could} be detectable if the
$r$-mode instability scenario holds.
We assume $A=3/2$ as above.
We plot only pulsars with known spin-downs, corrected for kinematic effects when
those are known.
Filled diamonds indicate $h_0$ due to $\alpha = 10^{-3}$ for pulsars where that
is allowed by the spin-down.
Hollow diamonds indicate the spin-down limit for pulsars where that means
$\alpha < 10^{-3}.$
The former pulsars are generally young and the latter are old.
Sensitivity curves in Fig.~\ref{f:rmodes} assume a sensitivity depth of
100\,Hz$^{-1/2},$ comparable to~\cite{Rajbhandari2021}, in the middle of the
range for $r$-mode searches which is about
75--130\,Hz$^{-1/2}$~\cite{Wette2023}.

In Table~\ref{tab:rmodes} we list the top few pulsars from Fig.~\ref{f:rmodes},
including \ac{GW} frequencies and frequency derivatives and distances.
For most of these pulsars $h_0$ is given by the spin-down limit, which
corresponds to $\alpha$ of order $10^{-5}$--$10^{-6}.$
The two at lowest frequency have much higher spin-down limits and $\alpha$ is
taken to be $10^{-3}.$
This list generally consists of nearby millisecond pulsars, often discovered in
recent years.
It has a lot of overlap with Table~\ref{tab:msps}, and like that Table most of
its pulsars have not been searched.
One of the exceptions is J0437\textminus4715, one of the nearest millisecond
pulsars which was searched recently \cite{O4aPulsars}.

\begin{table}
\begin{tabular}{lcrcr}
\hline
\hline
\multicolumn{1}{c}{JName} &
\multicolumn{1}{c}{$h_0$} &
\multicolumn{1}{c}{$f$} &
\multicolumn{1}{c}{$-\dot{f}$} &
\multicolumn{1}{c}{$D$}
\\
&
\multicolumn{1}{c}{} &
\multicolumn{1}{c}{(Hz)} &
\multicolumn{1}{c}{$\left( \mathrm{Hz}\, \mathrm{s}^{-1} \right)$} &
\multicolumn{1}{c}{(pc)}
\\
\hline
J1720\textminus0533 & $8.90\times10^{-27}$ & 459 & $1.15\times10^{-15}$ & 191
\\
J0955\textminus3947 & $9.53\times10^{-27}$ & 742 & $1.39\times10^{-14}$ & 488
\\
J1737\textminus0811 & $7.06\times10^{-27}$ & 359 & $6.57\times10^{-16}$ & 206
\\
J0711\textminus6830 & $7.01\times10^{-27}$ & 273 & $7.42\times10^{-16}$ & 106
\\
J0307+7443 & $6.51\times10^{-27}$ & 475	& $2.60\times10^{-15}$ & 386
\\
J0605+3757 & $6.58\times10^{-27}$ & 550	& $9.51\times10^{-16}$ & 215
\\
J0838\textminus2827 & $4.30\times10^{-27}$ & 415 & $1.13\times10^{-15}$ & 413
\\
J0437\textminus4715 & $4.11\times10^{-27}$ & 261 & $2.59\times10^{-15}$ & 157
\\
J1012\textminus4235 & $4.16\times10^{-27}$ & 484 & $1.00\times10^{-15}$ & 372
\\
\hline
\end{tabular}
\caption{
\label{tab:rmodes}
Like Table~\protect\ref{tab:msps} but for the top nine pulsars from
Fig.~\protect\ref{f:rmodes} with $r$-mode frequencies ($A=3/2$).
For most of these pulsars $\alpha$ is the spin-down limit $\alpha_\mathrm{sd},$
generally of order $10^{-5}$--$10^{-6}.$
For J0711\textminus6830 and J0437\textminus4715, $\alpha$ is $10^{-3}$ and well
below the spin-down limit.
}
\end{table}

\section{Directed searches for non-pulsing neutron stars}
\label{s:directed}

Many neutron stars observed via \ac{EM} waves from radio to x-rays are not
observed to pulse.
In many cases the pulsar beam simply does not align right to be seen from Earth.
In rapidly accreting neutron stars pulsations might be hidden by the accreting
material.

Astrophysical reasons and the mechanics of directed continuous \ac{GW} searches
naturally divide these into two types, the ``Cas A'' type searches for
non-accreting objects~\cite{S5CasA} and ``Sco X-1'' type searches for accreting
objects~\cite{FirstScoX1}.
First we describe how the Cas A type searches \textit{could} detect \acp{GW},
then finish with how the Sco X-1 type searches \textit{should} detect \acp{GW}.

\subsection{Non-accreting}

\begin{figure}
\includegraphics[width=0.45\textwidth]{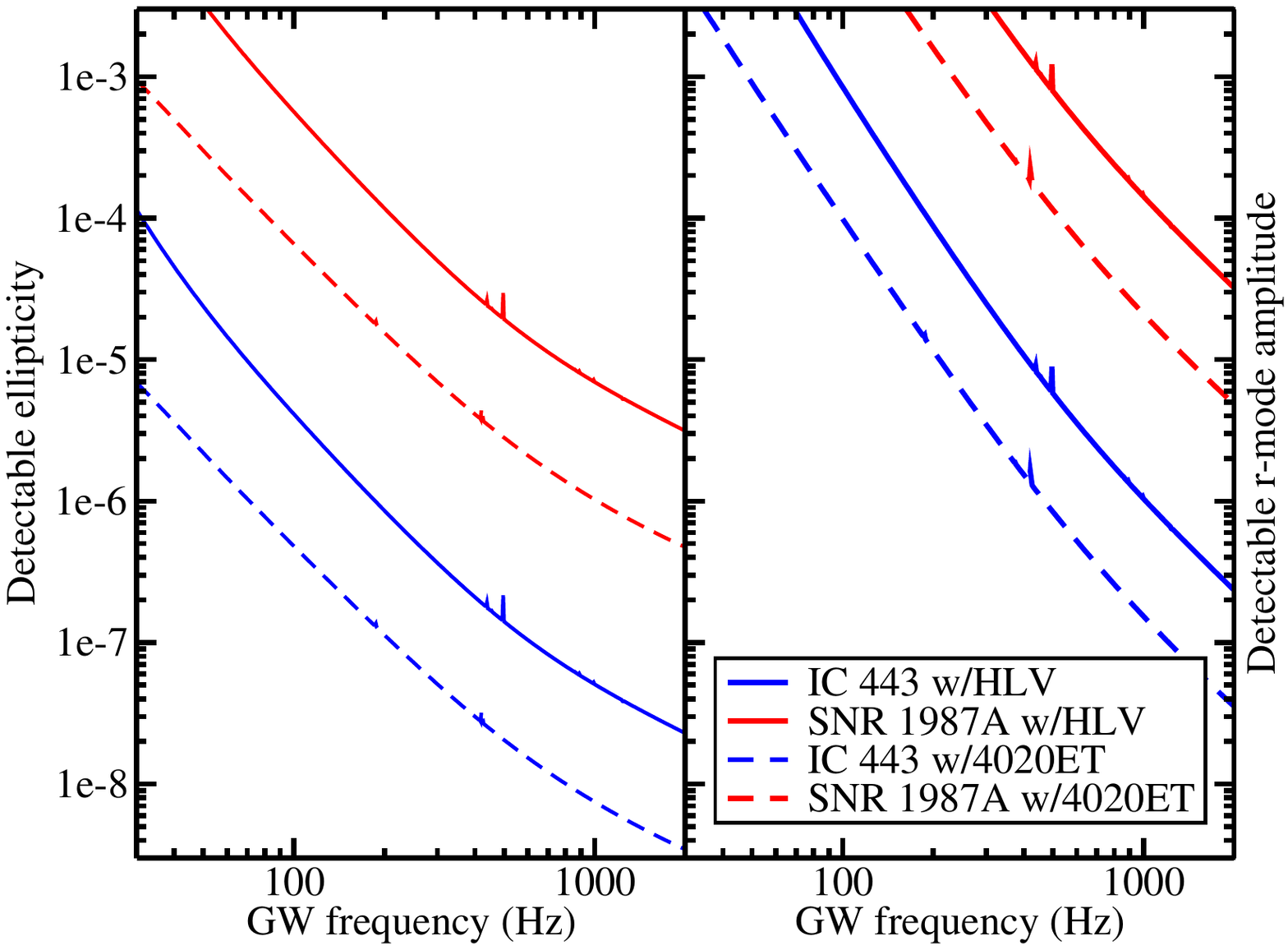}
\caption{
\label{f:snrs}
Detectable ellipticity (left side) and $r$-mode amplitude (right side) for a
tricky target (SNR\,1987A) and an easier target (IC\,443) for a modestly
improved network (HLV operating in O5) and for a greatly improved network
(4020ET).
Sensitivity depths are 20\,Hz$^{-1/2}$ for 1987A and 80\,Hz$^{-1/2}$ for
IC\,443.
The variation between targets is greater than the variation between networks.
For standard theoretical predictions of maximum ellipticity and $r$-mode
amplitude, an easy target is detectable above tens of Hz and a tricky target is
detectable above hundreds of Hz.
}
\end{figure}

Most of the interesting targets for directed searches are non-accreting.
These include central compact objects in supernova remnants, small pulsar wind
nebulae, and a few young supernova remnants or SNRs \cite{S6SNRs}.
Such objects tend to be fairly young, tens of thousands of years at most.
That is good because an analog of the spin-down limit based on the age of the
object \cite{Wette2008},
\begin{equation}
h_0^\mathrm{age} = 1.26 \times 10^{-24} \left( \frac{\mbox{3.30\,kpc}} {D}
\right) \left( \frac{\mbox{0.3\,kyr}} {a} \right)^{1/2},
\end{equation}
yields better milestones for younger objects.
The above equation is for \ac{GW} emission due to ellipticity.
For $r$-modes it is multiplied by a constant which is nearly
unity~\cite{Owen2010}.
It is analogous to the spin-down limit because it assumes that the spin-down has
been dominated by \ac{GW} emission from birth.
Youth is also good from a theory point of view, since $r$-mode amplitudes might
be higher \cite{Bondarescu2009} and perhaps the asymmetry of the supernova that
gave birth to the neutron star left an asymmetric mass distribution.

Searches of this type cover broad frequency bands since the lack of pulses means
the star's spin frequency and frequency evolution are unknown.
The \ac{GW} frequency is generally modeled as nearly constant, and the range of
parameters to search is greater for younger objects.
Thus searches for young objects cost more and are somewhat less sensitive.
The neutron star in SNR~1987A is young (less than 40\,yr old) and distant
(51.4\,kpc), and thus presents a relatively difficult and less sensitive search
which was only done recently for a physical parameter space~\cite{Owen2022,
Owen2024}.
IC\,443 (also known as SNR\,G189.1+3.0) is relatively old (we take 20\,kyr) and
nearby (1.5\,kpc), and is thus relatively easy and more sensitive---indeed, it
has been searched since~\cite{S6SNRs}.
We use these two examples to bracket the range of possibilities in our plots.

The plots in Figure~\ref{f:snrs} estimate sensitivities for these two targets
with the HLV and 4020ET networks.
Since searches for both of these have already reached $h_0^\mathrm{age},$ a plot
of $h_0$ vs.\ frequency like the others is not very informative---these sources,
and more, will be detectable at $h_0^\mathrm{age}$ over wide frequency bands by
all future networks.
It is more informative to look at detectable values of $\epsilon$ and $\alpha.$
All curves plot the sensitivity in terms of $\epsilon(f)$ or $\alpha(f)$ derived
from $h_0(f)$ using Eq.~(\ref{h0eps}) or~(\ref{h0alpha}) respectively.
For each network one curve shows a sensitivity depth of 20\,Hz$^{-1/2}$
(SNR\,1987A) and one shows 80\,Hz$^{-1/2}$ (IC~443), both based on previous
searches~\cite{Wette2023}.
Standard predictions of maximum $\epsilon \sim 10^{-5}$ and $\alpha \sim
10^{-3}$ for normal neutron stars indicate that easy targets are detectable
above 100\,Hz in the near future and tens of Hz with Cosmic Explorer.

\subsection{Accreting}

\begin{figure}
\includegraphics[width=0.45\textwidth]{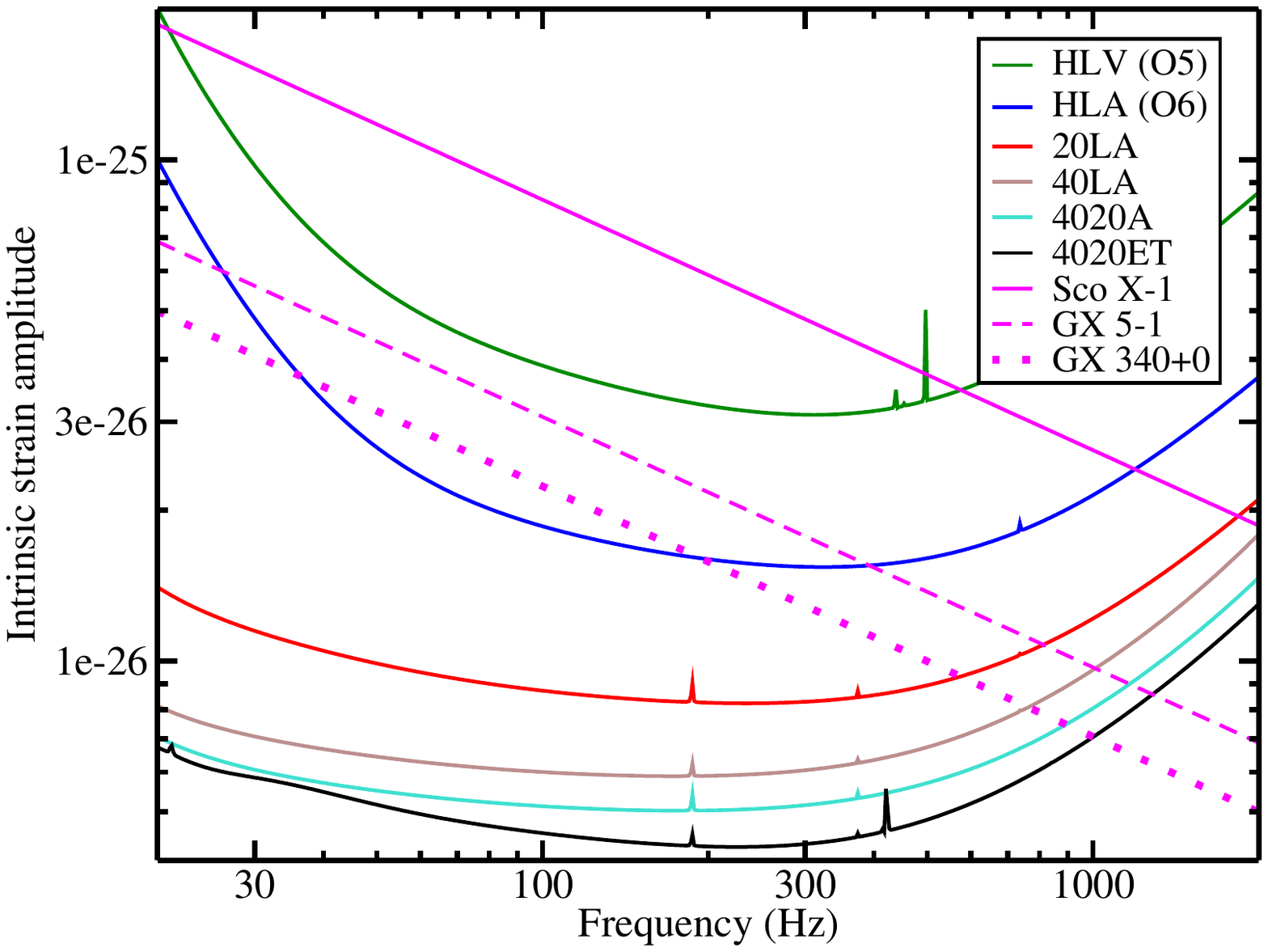}
\caption{
\label{f:lmxbs}
\acp{LMXB} that \textit{should} be detected by various networks if they emit at
the right frequency.
Slanted lines show the torque balance value of intrinsic strain assuming mass
quadrupole emission and fiducial neutron star parameters for the top few
binaries ranked by x-ray flux.
Solid curves show sensitivities for various interferometer networks assuming a
sensitivity depth of 60\,Hz$^{-1/2}.$
}
\end{figure}

We finish with another class of objects which \textit{should} be detected, the
\acp{LMXB} or rapidly accreting neutron stars.
Accretion onto neutron stars is known to be asymmetric, since in some cases
pulses due to rotating hot spots are seen.
The surface magnetic field should funnel accreted plasma toward the magnetic
poles and should to some extent support mountains \cite{Melatos2005}.
Hot spots extending down into the crust should lead to buried mass asymmetries
due to electron capture rates \cite{Bildsten1998}.
Centrifugal force due to accretion spin-up might globally break the
crust~\cite{Morales2024}.
$R$-modes could be unstable in rapidly rotating stars kept warm by accretion
\cite{Bildsten1998, Andersson1999b}.
Regardless of the details of producing the asymmetry, there is an old argument
\cite{Papaloizou1978b, Wagoner1984, Bildsten1998} that accretion onto neutron
stars with weak surface magnetic fields should spin them up until \ac{GW}
spin-down balances accretion torque spinning them up.
There is also evidence~\cite[most recently]{Gittins2019} that the spin periods
observed in these objects are consistent with \ac{GW} emission playing an
important role in them, even including the effects of the popular Ghosh-Lamb
(non-GW) accretion model~\cite{Cikintoglu2023}.

If the torque balance argument is correct, it results in a \ac{GW} intrinsic
strain~\cite{Wagoner1984, Watts2008, O3ScoX1}
\begin{eqnarray}
\label{torque}
h_0 &=& 9.49\times10^{-26} \left( \frac{F_x} {10^{-8}\,\mathrm{erg\,cm^{-2}\,s^{-1}}
}
\right)^{1/2} \left( \frac{P} {\mathrm{1\,s}} \right)^{1/2}
\nonumber\\
& & \times \left( \frac{R} {\mbox{10\,km}} \right)^{3/4} \left( \frac{M}
{1.4\,M_\odot} \right)^{-1/4}
\end{eqnarray}
assuming that accretion occurs on the surface of the star and not at the
Alfv\'en radius (in which case $h_0$ may be roughly 10\%
higher~\cite{Zhang2021}).
Here $F_x$ is the x-ray flux observed at Earth, so this argument favors the
apparently brightest x-ray sources in the sky such as Sco~X-1.
For $r$-mode emission this equation (in terms of spin period) remains the
same~\cite{Owen2010}, but it is different from ellipticity when converted to
\ac{GW} frequency $f.$

There are caveats, summarized e.g.\ in~\citet{Watts2016}.
Some neutron stars in \acp{LMXB} pulse, at least occasionally, and their spins
do not pile up near the upper limit.
Observations could be subject to various selection effects.
Perhaps interaction of accreting plasma with the star's magnetic field is enough
to brake stars without \ac{GW} emission, or at least many stars.
Stars with $r$-modes might emit significant \acp{GW} for a small fraction of the
time~\cite{Levin1999, Spruit1999}.

Also, these \ac{GW} searches are more difficult than those for known pulsars.
The best sources (brightest in x-rays) do not have timing solutions because no
pulses are observed.
X-ray fluxes of these sources fluctuate stochastically, indicating that
accretion torque also fluctuates and thus the spin frequency may wander
unpredictably on timescales of tens of days.
Orbital parameters are sometimes poorly constrained.
Still, these are attractive targets.
They have been searched since the initial detector era~\cite{O1ScoX1}, starting
with Sco~X-1 which is the best due to its high x-ray flux.

In Fig.~\ref{f:lmxbs} we plot torque balance limits for several of the best
(highest $F_x$) targets compared to projected sensitivities of various detector
networks.
Torque balance curves (the straight lines) are taken from Eq.~(\ref{torque})
with fiducial parameters, assuming average x-ray fluxes from
Ref.~\cite{Watts2008}.
Sensitivity curves assume a depth of 60\,Hz$^{-1/2},$ comparable to the best
achieved before this paper was submitted~\cite{Wette2023}.
(After this paper was submitted, the result from the first part of O4
\cite{O4aScoX1} appeared on arXiv with a depth of 70--75\,Hz$^{-1/2}.$
That and the previous result \cite{O3ScoX1} already beat the torque balance
limit over narrow frequency ranges under certain assumptions.)
This plot shows that, depending on the \ac{GW} frequency and assuming torque
balance, Sco~X-1 is likely to be detected by present generation facilities and
almost certain to be detected by next generation facilities.
(Although see \cite{Pagliaro2025} for a more complicated recent model suggesting
lower probabilities.)
With next generation detectors, several other systems likely become detectable
if the torque balance argument holds.
At that point, if nothing is detected it is very interesting because \ac{GW}
torque balance is likely ruled out as explaining the spins of these systems.

\section{Summary and conclusions}
\label{s:conclusion}

We have shown that there is cause to be optimistic that continuous gravitational
waves will be detected with modest improvements to interferometer networks in
the next few years, and that---if our current understanding of accreting neutron
stars and millisecond pulsars is correct---next generation interferometers such
as Cosmic Explorer and the Einstein Telescope are highly likely to detect many
such signals.
Even if nothing is detected in that era, it will spur major revisions of our
understanding of compact objects.
We have improved on previous estimates by giving detailed sensitivity curves
extrapolated from previous searches, with all those practical limitations of
those searches included, and by considering numerous specific targets currently
known from electromagnetic astronomy.
Improvements in data analysis techniques will improve sensitivities further.
Future astronomical discoveries from radio surveys (finding many millisecond
pulsars) to x-ray timing (potentially discovering the spin period of Sco\,X-1)
will also improve the sensitivity of gravitational wave searches.
The future of multi-messenger astrophysics is bright and is not limited to
compact binary coalescences.

\begin{acknowledgments}

This work was supported by NSF grants PHY-1912625 and PHY-2309305 to TTU,
PHY-2110460 and PHY-2409745 to RIT, and PHY-2450793 to UMBC.
We are grateful for helpful discussions with Alessandra Corsi and Michael
Kr\"amer about present and future radio observations, with Matthew Pitkin and
Graham Woan about millisecond pulsars, and with Anthony Pearce about sensitivity
in the early stages of this work.

\end{acknowledgments}

\bibliography{ce1}

\end{document}